\documentclass{PoS}
\usepackage{subfigure}

\usepackage{amsmath}

\def\babar{\mbox{\sl B\hspace{-0.4em} {\small\sl A}\hspace{-0.37em}
    \sl B\hspace{-0.4em} {\small\sl A\hspace {-0.02em}R}}}

\def\RD_value{3.51 \pm 0.35}
\def\yp_value{4.3 \pm 4.3}
\def\xp2_value{0.08 \pm 0.18}

\title{Observation of charm mixing at CDF}

\ShortTitle{Observation of charm mixing at CDF}

\author{\speaker{Paolo Maestro}\\ On behalf of the CDF collaboration\\
        Department of Physics, University of Siena and INFN, via Roma 56, 53100 Siena, Italy\\
        E-mail: \email{paolo.maestro@pi.infn.it}}

\abstract{    
  We report on the observation of $D^0$\---$\bar{D}^0$ oscillations 
  by measuring
  the time-dependent ratio of yields for the
  rare decay $D^0 \rightarrow K^+\pi^-$ to the favored decay
  $D^0 \rightarrow K^-\pi^+$ at the Collider Detector at Fermilab (CDF).
  Using 9.6 fb$^{-1}$ of integrated luminosity of $\sqrt{s}$ = 1.96 TeV $p$$\bar{p}$ collisions
  recorded in the full CDF Run II, 
  the signals of
  $7.6 \times 10^6$
  $D^0\rightarrow K^-\pi^+$ and
  $33 \times 10^3$ 
  $D^0\rightarrow K^+\pi^-$ decays are reconstructed in $D^{*}$-tagged events, with proper decay times
  between 0.75 and 10 mean $D^0$ lifetimes.
  We measure the mixing parameters   $x'^2 = (0.08 \pm 0.18)\times 10^{-3}$,
  $y' = (4.3 \pm 4.3) \times 10^{-3}$, and
  $R_D = (3.51 \pm 0.35) \times 10^{-3}$.
  Our results are consistent with standard model expectations and  similar results from proton-proton collisions
  and exclude 
  the no-mixing hypothesis 
  with a significance equivalent to 6.1 standard deviations. 
}

\FullConference{14th International Conference on B-Physics at Hadron Machines,\\
		April 8-12, 2013\\
		Bologna, Italy }

\begin{document}
\section{Introduction}
Neutral mesons can oscillate into their antiparticles 
because they are produced in flavor eigenstates which are different 
from eigenstates with defined mass and lifetime.
This quantum-mechanical oscillation is referred as mixing and 
 can be characterized by the parameters $x = \Delta m / \Gamma$
and $y = \Delta \Gamma / 2 \Gamma$, where $\Delta m$ is the mass
difference, $\Delta \Gamma$ is the decay width difference, and $\Gamma$ is
the mean decay width of the mass eigenstates.
The process  is well established for $K^0$, $B^0$, and
$B^0_s$ mesons \cite{ref:PDG} and its study
provides important information about electroweak interactions and the
Cabibbo-Kobayashi-Maskawa (CKM) matrix, as well as the virtual particles that
are exchanged in the mixing process itself.
Evidence of $D^0$\---$\bar{D}^0$ mixing was reported in recent years by the experiments
Belle \cite{ref:Belle_mixing}, Babar \cite{ref:BaBar_mixing, ref:BaBar_mixing_lifetime_diff} and CDF \cite{ref:CDF_mixing}
and it was  observed only in 2012 by LHCb \cite{ref:LHCb}.
Even if standard model (SM) calculations of the $D^0$\---$\bar{D}^0$ mixing rate are affected by significant theoretical uncertainties, 
this process is expected to be much slower (i.e. $|x|$, $|y| \le 10^{-3}$) than the  $B$ and $K$ oscillations.
However NP particles could  enhance the mixing rate, 
thus providing indirect evidence for physics beyond the SM \cite{ref:mixreview, ref:GHPP-2007}.
Then, it is of great interest to establish conclusively
$D^0$\---$\bar{D}^0$ mixing in a specific decay channel and improve the precision of 
the measurement of the mixing parameters.
\subsection{Charm mixing in the $D^0 \to K^+\pi^-$ channel}
Charm mixing can be searched by measuring the 
time dependence of the rate of the rare $D^0 \to K^+\pi^-$  decay (including its charge-conjugate). 
This decay can arise from the oscillation of a $D^0$ state to a $\bar{D}^0$ state, followed by
a Cabibbo-favored (CF) $D^0 \to K^-\pi^+$ decay, or from a doubly-Cabibbo suppressed (DCS) $D^0$ decay.
Under the assumption that CP is conserved and the mixing parameters are small ($|x|$, $|y|\ll 1$),
the ratio $R$ of $D^0 \to K^+\pi^-$ to $D^0 \to K^-\pi^+$ decay rates can be
approximated by \cite{ref:PDG}
\begin{equation}
R(t) = R_D + \sqrt{R_D} \, {y}^{\prime} \, t + \frac{x^{\prime 2} + y^{\prime 2}}{4} \, t^2 
\label{eqn:R(t)}
\end{equation}
where $t$ is the proper decay time 
expressed in units of mean $D^0$ lifetime.
$R_D$ is the DCS decay rate
relative to the CF rate, while the parameters $x^{\prime}$ and $y^{\prime}$ are linear combinations of $x$ and
$y$ according to the relations
$x^{\prime} = y ~{\rm sin} ~\delta_{K\pi} + x ~{\rm cos}~\delta_{K\pi}$   and 
$y^{\prime} = y ~{\rm cos} ~\delta_{K\pi} - x ~{\rm sin} ~\delta_{K\pi}$,  
where $\delta_{K\pi}$ is the strong interaction phase difference between the
DCS and CF amplitudes.  In the absence of mixing, $x' = y' = 0$ and $R(t) = R_D$. \\ 
The experimental method to identify the flavor of the charmed meson at production
exploits the strong-interaction decays
 $D^{*+} \to \pi^+ D^0$, $D^{*-} \to \pi^- \bar{D}^0$. 
The relative charges of the soft (low-momentum) tagging pion from $D^*$ decay and 
the pion from $D^0$ decay determine whether the
decay chain is right-sign (RS, like charge) or wrong-sign (WS, opposite charge).
RS processes include mainly CF decays, while 
DCS and mixing decays contribute to WS processes.
\section{Analysis}
Our measurement uses  the full data set corresponding to an integrated luminosity of $9.6$~fb$^{-1}$ recorded by the
CDF~II detector at the Tevatron in $p\bar{p}$ collisions at $\sqrt{s}=1.96$~TeV.
We reconstruct the 
WS $D^{*+} \to \pi^+ D^0 ( \to K^+ \pi^-)$, and
the RS  $D^{*+} \to \pi^+ D^0 (\to K^- \pi^+)$  decay chains 
and measure the time dependence of their rates ratio.
The components of the CDF~II detector  most relevant for this analysis are the
multi-wire drift chamber (COT) and the silicon
microstrip vertex detector located inside a solenoid, which provides a 1.4~T magnetic field \cite{ref:CDF}. 
\subsection{Data selection}
The events for this analysis are selected online by a trigger system
\cite{ref:CDF-RB} which identifies pairs of oppositely charged particle tracks
from a decay vertex detached by at least 200 $\mu$m from the beamline.
In the off-line analysis, 
the  tracks  satisfying the trigger requirements
are considered with both $K^-\pi^+$ and $\pi^-K^+$
interpretations in order to reconstruct $D^0$ candidates.
Minimal requirements on the momenta
and impact parameters of the tracks and the displacement of the reconstructed $D^0$ decay vertex are imposed.
A low-momentum tagging pion track 
is combined with the $D^0$ candidate to  form a $D^*$ candidate.
To reduce the contribution
of $D^*$ mesons produced from $b$-hadron decays, $D^0$ candidate are required to have an impact parameter $d_0<60~\mu$m.
RS $D^0$ decays incorrectly reconstructed as WS decays, because
the kaon and pion assignments are mistakenly interchanged, represent
a large background to the WS signal. 
Two selection cuts have been applied to reduce this background. 
WS candidates with RS $K\pi$ invariant mass reconstructed within 20
MeV/$c^2$ of the known $D^0$ mass are removed.
This cut retains 78\% of the WS signal, 
and rejects 96.5\% of the RS $D^0$ decays with incorrect mass assignment. 
A second cut  exploits the K/$\pi$ separation based on the measurements of the ionization energy loss in the COT \cite{ref:CDF-RB}. 
The combination of these two cuts greatly reduces the mis-assigned RS background
improving the WS signal over background ratio  by a factor $\sim100$.
\subsection{Signal extraction}
\begin{figure}
  \centering
\subfigure[]{\label{fig:Dmrs}
  \includegraphics[width=.45\textwidth]{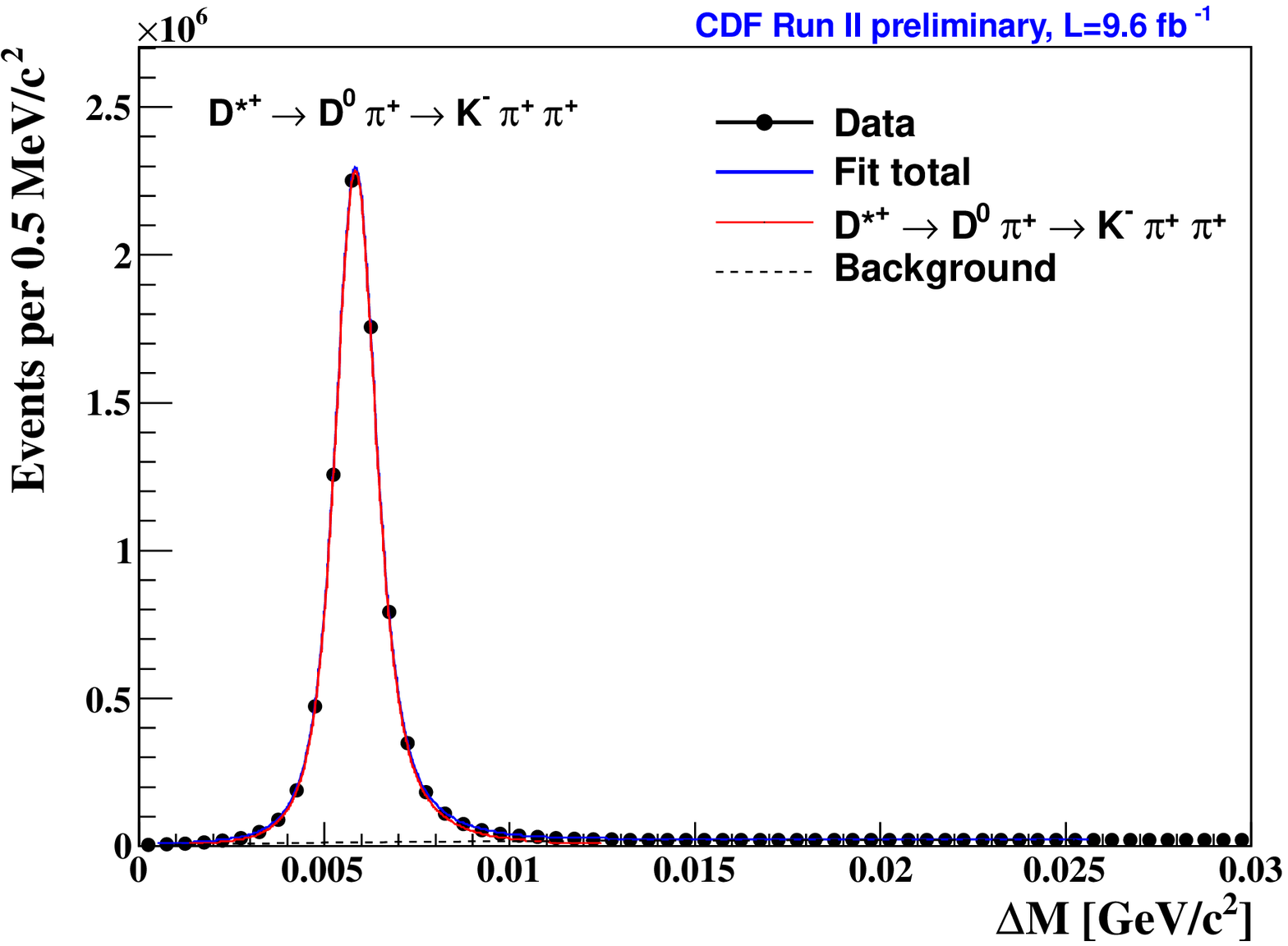}
}
\subfigure[]{\label{fig:Dmws}
  \includegraphics[width=.45\textwidth]{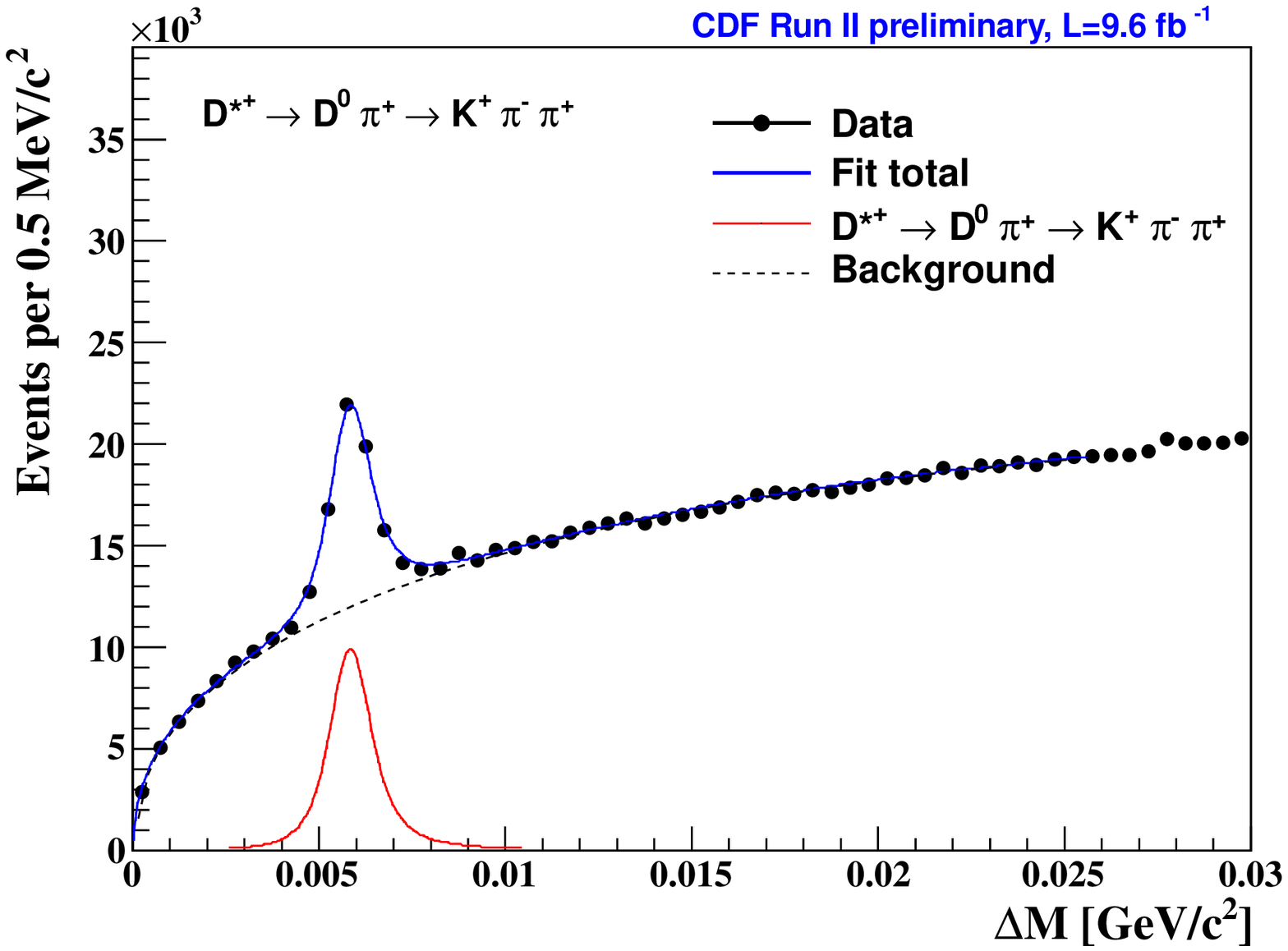}
}
\caption{
 Time-integrated $\Delta M$ distribution for (a) RS  and 
 (b) WS decays, with fit curves superimposed.  
}
\label{fig:WS_mass_diff}
\end{figure}

The reconstructed RS and WS condidates are classified into 20 intervals of proper decay time $t$ 
which is determined  (normalized to the mean $D^0$ lifetime $\tau=$ 410.1~fs) as $t = m_{D^0} L_{xy} / (p_T \tau)$, where 
$m_{D^0} = 1.8648$~GeV/$c^2$ is the known $D^0$ mass \cite{ref:PDG}, 
$L_{xy}$ is the transverse $D^0$ decay length, 
and $p_T$ its transverse  momentum.
$D^0$ candidates in each time bin are further  
divided into 60 bins of mass difference $\Delta M$ $\equiv$
$M(K\pi\pi) - M(K\pi) -M(\pi)$, with equal size 0.5 MeV/c$^2$.
For each of the resulting 1200 WS and 1200 RS bins, 
the $D^0$ signal yield is determined by fitting the
corresponding binned distribution of the K$\pi$ invariant mass $M_{K\pi}$. 
The signal shape is modeled by the sum of two Gaussian functions with a low-mass tail,
and the combinatoric background by an exponential function. 
A Gaussian term is included in the WS fit to model the residual  background from misidentified RS decays,
with shape determined from the data. 
The $D^*$ signal for each
time bin is determined from a $\chi^2$ fit of the $D^0$ signal
yield versus $\Delta M$.  The signal shape is modeled by a
double-Gaussian and an asymmetric tail function, the background shape
by the product of a power-law and an exponential function.  
The amplitudes of the signal and background and 
the background shape parameters
are determined independently for all $M_{K\pi}$ and $\Delta M$ fits, 
while the signal shape is fixed to the RS time-integrated shape. 
The $D^*$ fit procedure for the time-integrated  $\Delta M$
distributions is shown in Fig. \ref{fig:WS_mass_diff}. The fitted RS and WS
signal yields are about $7.6\times10^{6}$ and $33\times10^{3}$, respectively.
\subsection{WS/RS yield ratio}
\begin{figure}[h]
\begin{center}
\includegraphics[width=.45\textwidth]{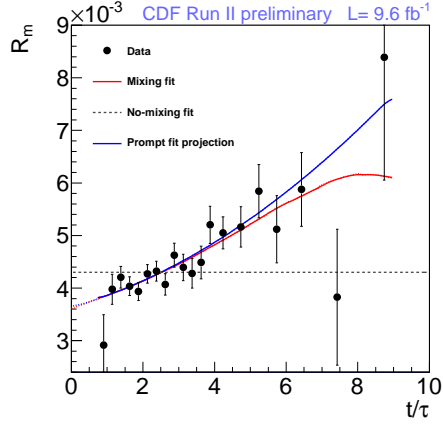}
\caption{
Measured ratio of WS to RS signal yields as 
a function of normalized proper decay time.}
\label{fig:ratio_data}
\end{center}
\end{figure}

\begin{figure}
  \centering
\subfigure[]{\label{fig:ip_dista}
  \includegraphics[width=.43\textwidth]{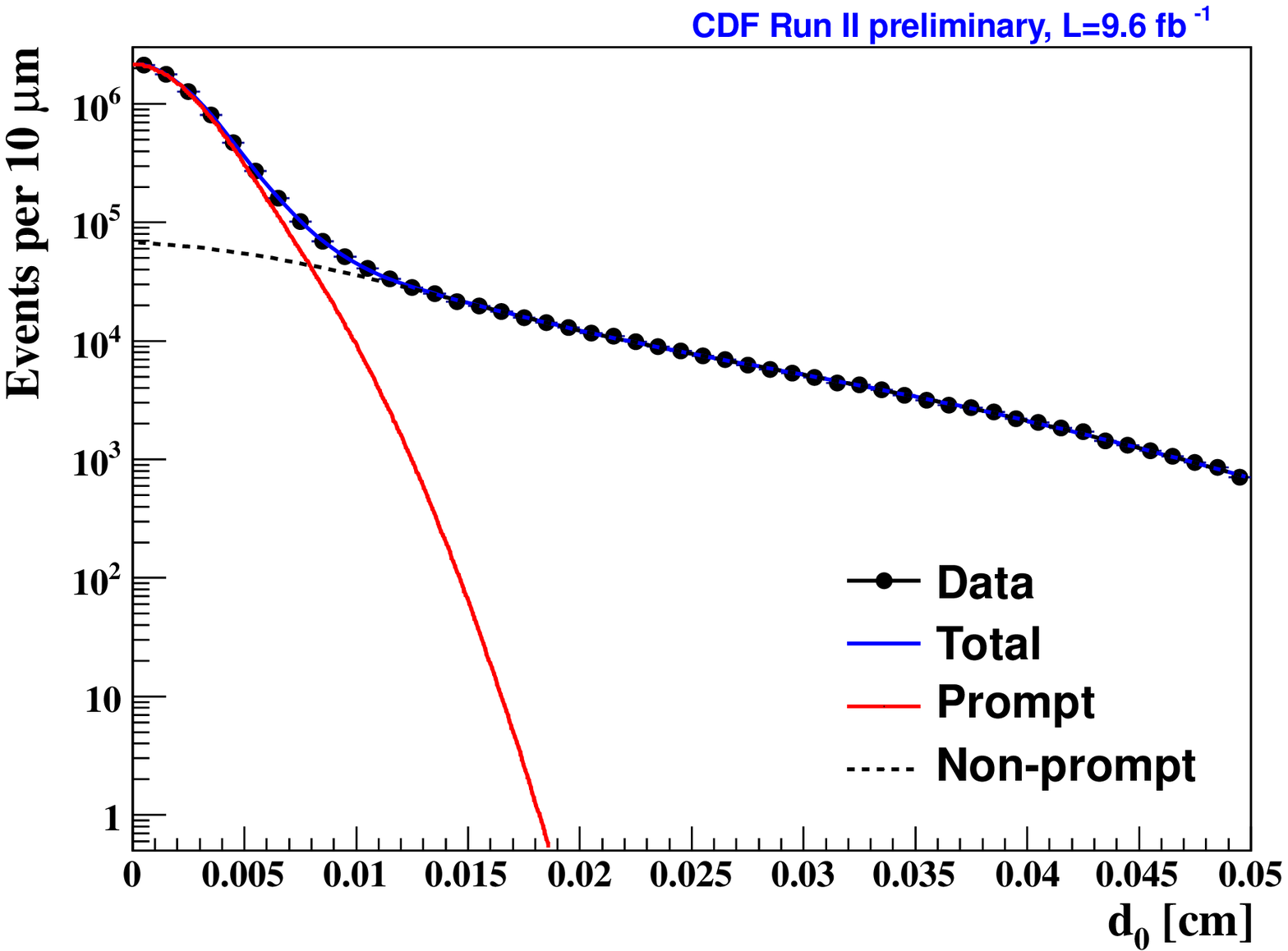}
}
\subfigure[]{\label{fig:ip_distb}
  \includegraphics[width=.45\textwidth]{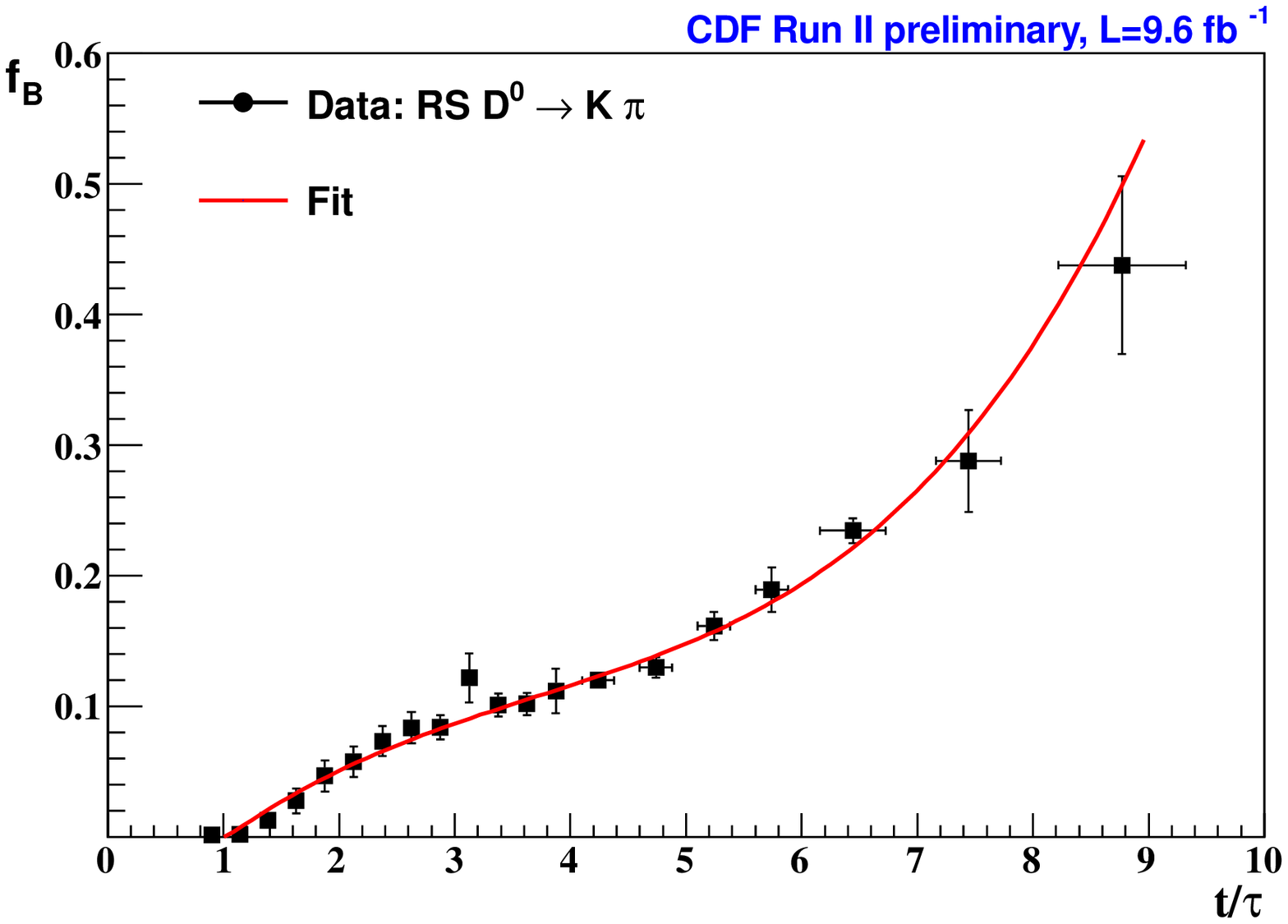}
}
\caption{
(a) Distribution of transverse impact parameter $d_0$ for 
RS $D^0$ candidates for all the time bins.
(b) Fraction of non-prompt RS $D^*$ decays as a function of proper decay time. 
}
\label{fig:ip_dist}
\end{figure}

The measured ratio $R_m$ of WS to RS signal yields in the 20 time intervarls
is shown in Fig.~\ref{fig:ratio_data}. 
Since the measured yields include the contribution of $D^*$ mesons produced from $b$-hadron decays, 
the time dependence of $R_m$ is different from that of the WS/RS ratio of prompt decays $R$
(Eq.~\ref{eqn:R(t)}). 
The expected value of $R_m$ in a given time bin
can be factorized as the product of $R$ by a correction factor
due to the non-prompt production
\begin{equation}
R_m^{pred}(t) 
= R(t) \left[ 1 + f_B(t) \left( \frac{R_B(t)}{R(t)} - 1 \right)\right] 
\label{eqn:R_expected}
\end{equation}
where  $f_B(t)$ is the fraction of non-prompt RS $D^*$ decays  and 
$R_B(t)$ is the WS/RS ratio of non-prompt $D^*$ decays with measured decay time $t$. 
For non-prompt decays, the measured decay time is the sum of the decay times of the beauty particle parent and the $D^0$ daughter. 
The function $R_B(t)$ is calculated by weighting $R(t)$ with the decay-time distributions of non-prompt $D^0$ decays 
obtained from a full detector simulation.
The function $f_B(t)$ is determined from data 
by fitting the $d_0$ distributions of RS $D^*$ decays in each time bin. 
These are characterized by a peak at small $d_0$ due to the prompt component, and a broad distribution
extending to large $d_0$ due to the non-prompt component, as shown in Fig.~\ref{fig:ip_dista}.
Both the prompt and non-prompt components are  modeled with the sum of two Gaussians. 
The time dependence of $f_B$ in the region $d_0<60 \mu$m   
is parametrized by a 4-degree polynomial (Fig.~\ref{fig:ip_distb}).\\
The mixing parameters $R_D$, $y'$, and $x'^2$ 
are found by minimizing the $\chi^2$ function
\begin{equation}
  \chi^2 = \sum_{i=1}^{20} \left[ 
\frac{R_m(t_i) - R_m^{pred}(t_i)} {\sigma_i} 
\right]^2 + C_B + C_H
\label{eqn:chi2_Rm}
\end{equation}
where $\sigma_i$ is the uncertainty on $R_m(t_i)$
and  $C_B$ and $C_H$ are Gaussian constraints to the parameters describing $f_B(t)$ and $R_B(t)$, respectively.\\
We investigated extensively
systematic uncertainties due to a
number of possible sources including:
detector charged track asymmetries, 
uncertainties in the signal shapes used to fit $M_{K\pi}$ and $\Delta M$ distributions
and in the shape of non-prompt component used to fit the $d_0$ distributions,
background due to $D^+ \rightarrow K^- \pi^+ \pi^+$ and partially reconstructed charm decays, 
sensitivity of $R_B(t)$ on the simulated decay time distributions of non-prompt $D^0$.
All these effects were found to be small compared to the mixing parameter errors derived from the fit. 
\subsection{Result}
\begin{figure}
  \centering
\subfigure[]{\label{fig:contoura}
\includegraphics[width=.48\textwidth]{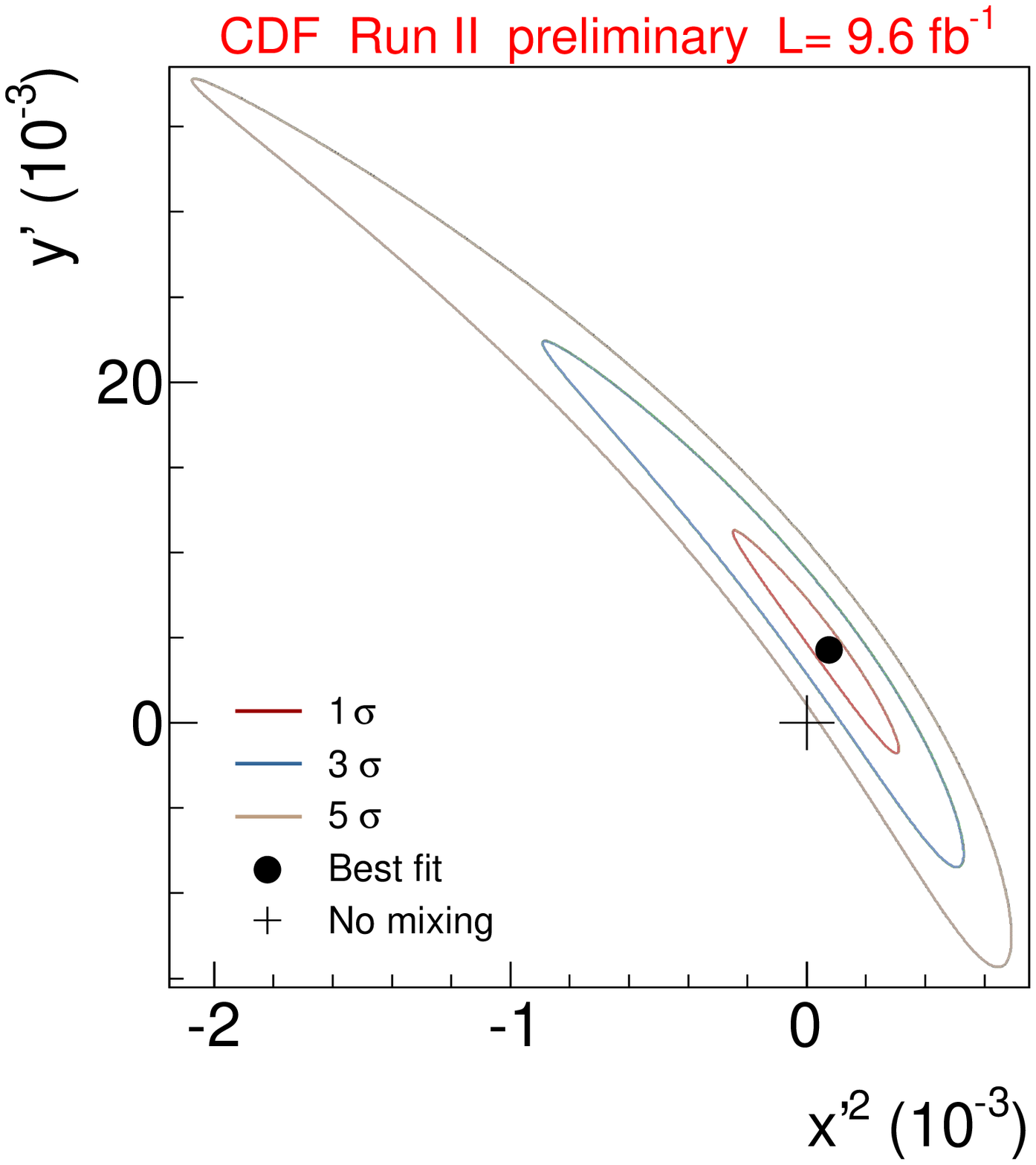}
}
\subfigure[]{\label{fig:contourb}
\includegraphics[width=.45\textwidth]{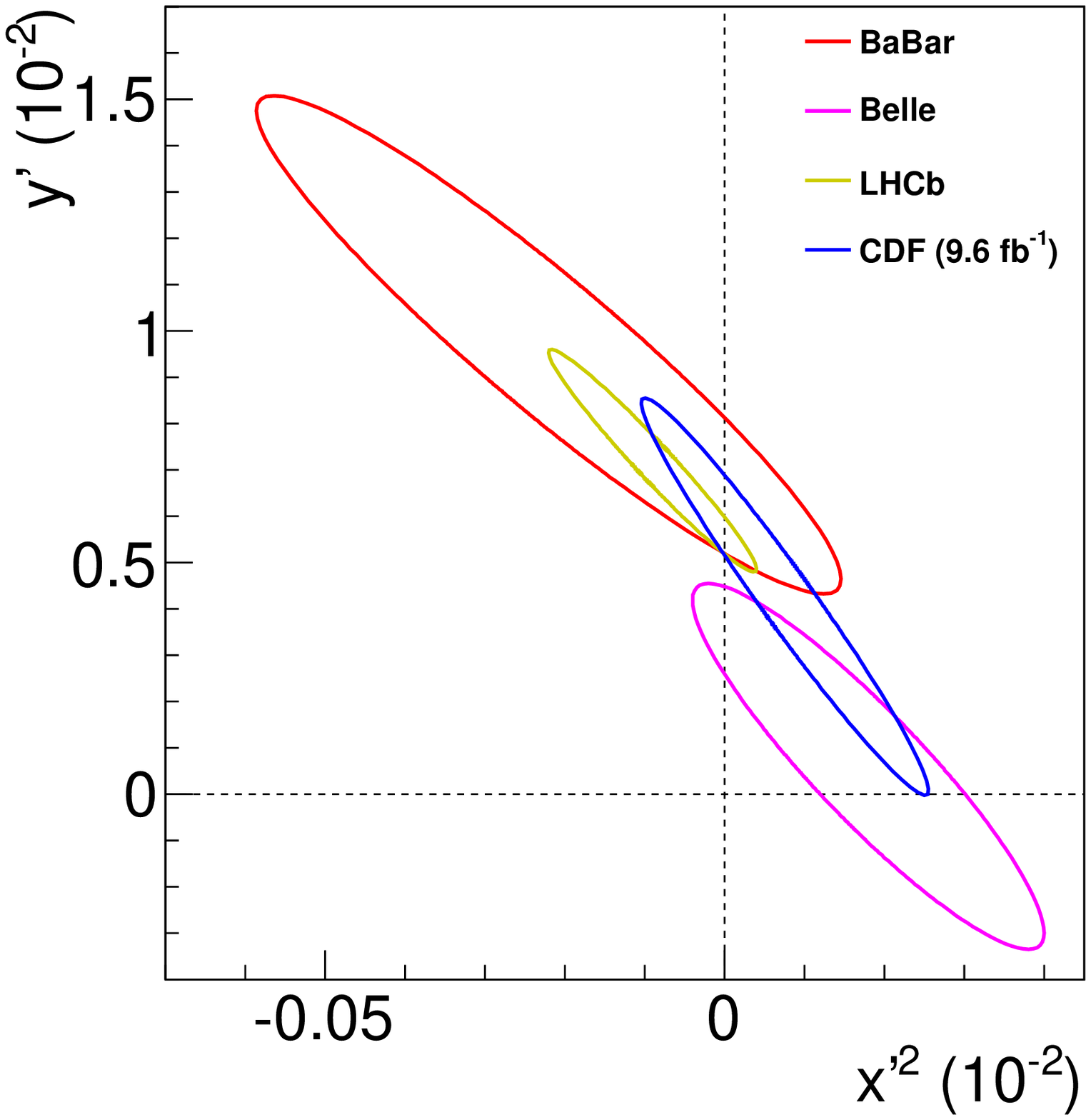}
}
\caption{
(a) Bayesian probability contours in $x'^2 - y'$ parameter space.
(b)  Comparison of 1$\sigma$ contours between CDF (this work),
Belle \cite{ref:Belle_mixing}, Babar \cite{ref:BaBar_mixing, ref:BaBar_mixing_lifetime_diff} and  LHCb \cite{ref:LHCb}.
}
\label{fig:contour}
\end{figure}

The fitted values of the
mixing parameters are reported in Table \ref{tab:cormatrix}. 
The function $R_m^{pred}(t)$ 
and the prompt component $R(t)$ as determined by the fit are shown in Fig.~\ref{fig:ratio_data}. 
They differ at large $t$ due to the effect of non-prompt $D^*$ production.  A fit
assuming no-mixing, i.e. $y' = x'^2 = 0$, is also shown and is clearly incompatible with the data.
\begin{table}
\begin{center}
\begin{tabular}{|c|c|c|c|cccc| } \hline
   Fit type & $   \chi^2$  /ndf &   Parameter  &    Fitted values &  & &    Correlation coefficient & \\
   &  &   &$   \times 10^{-3}$ & & $   R_D$ & $   y'$ & $   x'^2$\\ \hline 
   Mixing   &   16.91/17  & $   R_{D}$   & $   3.51\pm0.35$ &     &   1 &    -0.967 &
   0.900\\  
         &          &$   y'$   & $  4.3\pm4.3$ &     &  &    1 &    -0.975   \\
         &          &$   x'^{2}$ & $  0.08\pm0.18$ &    &  &   &     1 \\ \hline
   No-mixing&  58.75/19&$   R_D$ & $ 4.30\pm0.06$&  & & & \\   \hline
\end{tabular}
\end{center}
\caption{Mixing parameter results. The uncertainties include statistical and systematic components.}
\label{tab:cormatrix}
\end{table}
By calculating the Bayesian probability contours in the $x'^2$-$y$ parameter space
(Fig.~\ref{fig:contoura}), we exclude the no-mixing hypothesis at the level of 6.1 Gaussian standard deviations.
Our results are consistent with SM calculations \cite{ref:mixreview} and measurements from other experiments, 
as shown by comparing the 1$\sigma$ $x'^2$-$y$ contours in Fig.~\ref{fig:contourb}, 
and have similar precision to the recent LHCb observation \cite{ref:LHCb}.
\section{Conclusion}
We observe  $D^0$\---$\bar{D}^0$ mixing with a significance equivalent to 6.1$\sigma$, by measuring
the decay-time-dependence of the ratio of yields for the suppressed $D^0 \rightarrow K^+\pi^-$
to the favored  $D^0 \rightarrow K^-\pi^+$ decays using the full CDF data set. 
We measure the mixing parameters to be $R_D = (\RD_value) \times 10^{-3}$,
$y' = (\yp_value) \times 10^{-3}$, and $x'^2 = (\xp2_value)\times 10^{-3}$.
Our results are consistent with SM predictions and similar measurements from other experiments
and substantially improve global knowledge of the charm mixing parameters.


\begin{thebibliography}{99}
\bibitem{ref:PDG}
J. Beringer {\it et al.} (Particle Data Group), 
Phys.\ Rev.\ D {\bf 86}, 010001 (2012)
%
\bibitem{ref:Belle_mixing}
M.~Stari\v{c} {\it et al.} (Belle Collaboration),  
Phys.\ Rev.\ Lett. {\bf 98}, 211803 (2007)
%
\bibitem{ref:BaBar_mixing}
B.~Aubert {\it et al.} (\babar~Collaboration),
Phys.\ Rev.\ Lett. {\bf 98}, 211802 (2007)
%
\bibitem{ref:BaBar_mixing_lifetime_diff}
J.P. Lees {\it et al.} (\babar~Collaboration),
Phys. Rev. D {\bf 87}, 012004 (2013)
%
\bibitem{ref:CDF_mixing}
T.~Aaltonen {\it et al.} (CDF Collaboration),
Phys.\ Rev.\ Lett. {\bf 100}, 121802 (2008)
%
\bibitem{ref:LHCb}
LHCb Collaboration, R. Aaij {\it et al.} (LHCb Collaboration),
Phys. Rev. Lett. {\bf 110}, 101802 (2013)
%
\bibitem{ref:mixreview}
C.~A. Chavez, R.~F. Cowan, and W.~S. Lockman,
Int. J. Mod. Phys. A {\bf 27}, 1230019 (2012) 
%
%
\bibitem{ref:GHPP-2007}
E. Golowich, J. Hewett, S. Pakvasa, and A.~A.~Petrov,
Phys.\ Rev.\ D  {\bf 76}, 095009 (2007)
%
%
%
%
\bibitem{ref:CDF}
D.~Acosta {\it et al.} (CDF Collaboration),
Phys.\ Rev.\ D {\bf 71}, 032001 (2005)
%
\bibitem{ref:CDF-RB} 
A. Abulencia {\it et al.} (CDF Collaboration), Phys. Rev. D {\bf 74},
031109(R) (2006)

\end{thebibliography}
\end{document}